\documentstyle[aps,prd,preprint,epsfig]{revtex}

\begin{document}
\preprint{LPT-ORSAY 02-96}
\draft
\title{Brane world in a texture}

\author{Inyong Cho\footnote{Electronic address: 
Inyong.Cho@th.u-psud.fr}}
\address{Laboratoire de Physique Th\'eorique,
Universit\'e Paris-Sud, B\^atiment 210, F-91405   
Orsay CEDEX, France}

\date{\today}

\maketitle

\begin{abstract}
We study five dimensional brane physics 
induced by an O(2) texture formed in
one extra dimension.
The model contains two 3-branes of nonzero tension,
and the extra dimension is compact.
The symmetry-breaking scale of the texture controls
the particle hierarchy between the two branes.
The TeV-scale particles are confined to the negative-tension 
brane where the observer sees gravity as essentially 
four dimensional.
The effect of massive Kaluza-Klein gravitons is suppressed.
\end{abstract}

\vspace{0.5in}

\section{Introduction}
The idea of large extra dimensions has attracted
much attention since it was introduced~\cite{ADD}.
In particular, the Randall-Sundrum scenario
opened a distinct view of extra dimensions
by taking a nonfactorizable geometry 
into an account~\cite{RSI}.
The scenario contains two 3-branes with nonzero tensions.
One brane has a positive tension, and the other has
a negative tension.
The extra dimension is compact.
The bulk is filled with a negative cosmological constant.
The nonfactorizable geometry (the warp factor) of the extra dimension 
and the brane tensions
are determined solely by the bulk cosmological constant.

The warp factor is peaked on the positive-tension brane,
and decreases exponentially towards the negative-tension brane.
The effective-particle scale is also peaked on the
positive-tension brane, and decreases exponentially 
along the extra dimension due to the warp factor. 
Planck-scale particles are confined to the positive-tension
brane, and TeV-scale particles are in the negative-tension brane.

The localization of gravity in this scenario has been investigated
in Refs.~\cite{RSI,RSII,RSIII}.
The massless graviton which is responsible for 4D gravity
is localized on the positive-tension brane.
The amplitude of the zero-mode graviton is highest there,
and decreases away from that brane.
However, the observer on the TeV brane 
sees gravity as essentially four dimensional
even though the zero-mode amplitude is very small~\cite{RSIII}.
This is mainly because the effect of massive Kaluza-Klein 
gravitons is suppressed.

The scenario was posited by string theoretic motivations.
Subsequent work realized the four dimensional universe
more naturally, as a sub-manifold associated with topological 
defects formed by a
matter condensate in the extra dimensions~\cite{Defects}. 
The defect solution
determines the warped metric, and binds both gravitons and matter
fields to a four dimensional internal space.
Previous work has mainly focused on core defects
such as domain walls, cosmic strings, and monopoles.

In this paper, we investigate brane physics
induced by a global O(2) texture formed in an extra dimension~\cite{Texture}.
Textures are somewhat different from core defects.
Global textures are produced when a continuous global symmetry 
$G$ is broken to a group $H$.
The resulting homotopy group $\pi_D(G/H)$ is nontrivial,
and the vacuum manifold is $S^D$; for $G =\text{O}(N)$ models, 
$H=\text{O}(N-1)$ and $D=N-1$.
The mapping from the physical space to the vacuum manifold is
$R^D \to S^D$. 

After the symmetry is broken, the scalar field takes only
the vacuum-expectation value which is the broken-symmetry state.
There is no place in the physical space where the scalar
field remains in the unbroken-symmetry state.
This is the key difference between textures and core defects.

For a global O(2) texture, the mapping is $R^1\to S^1$.
We identify the physical space $R^1$ with the extra dimensional
space.
After the scalar field completes its winding in the vacuum manifold,
two spatial points on the boundaries of the physical space 
are mapped into the same point in the vacuum manifold.
These two points are identified, and the extra dimension
becomes compact.

The extra dimension is curved by the texture.
The strength of the curvature depends on the symmetry-breaking
scale of the texture.
This curved geometry of the extra dimension
sets the particle hierarchy as well as the Planck-mass
hierarchy in the extra dimension.
Thus, the role of the cosmological constant in the Randall-Sundrum 
scenario is replaced by the texture.

The model has two 3-branes of nonzero tension.
The one has a positive, and the other has a negative tension.
The particle-hierarchy scale between the two branes is determined
solely by the symmetry-breaking scale.
The Planck-scale particles are confined to the positive-tension
brane.
In order to have the particle scale be TeV on the negative-tension brane, 
the symmetry-breaking scale needs to be
super 5D Planckian scale.

Gravity on the TeV brane is essentially
four dimensional even though the amplitude of massless gravitons
is relatively small.
The higher dimensional effect of massive Kaluza-Klein gravitons
is strongly suppressed on this brane.
This result is similar to that of Ref.~\cite{RSIII}.

The structure of this paper is as following.
In Sec.~\ref{sec=model}, we present the model
and derive field equations.
In Sec.~\ref{sec=sol}, we obtain solutions
and discuss the particle hierarchy.
In Secs.~\ref{sec=zero} and \ref{sec=KK},
we study gravitons, and we conclude in Sec.~\ref{sec=con}.

\section{The model and field equations}\label{sec=model}
Let us consider an O(2) texture formed in one extra dimension.
The vacuum manifold is $S^1$, and it is mapped to $R^1$
which we set to be the extra dimensional space. 
We shall consider a texture of one winding number.
We assume that the extra dimension is compact with the 
$Z_2$ symmetry.
The two spatial points in the extra dimension, at which
the field takes the same value, are identified.
Then the orbifold is $S^1/Z_2$. 

The action describing the model of an O(2) texture 
and 3-branes in five dimensions is
\begin{equation}
{\cal S} = \int d^5x\sqrt{-g}
\left[ {{\cal R}\over 16\pi G_N}
-{1\over 2}\partial_\mu\Phi^a \partial^\mu\Phi^a
-{\lambda\over 4}(\Phi^a\Phi^a -\eta^2)^2 \right]
-\int_i d^4x\sqrt{-h} \sigma_i(\Phi^a)
\,,
\label{eq=action}
\end{equation}
where $\Phi^a$ is the scalar field with $a=1,2$,
$\eta$ is the symmetry-breaking scale,
$\sigma_i$ is the tension of the $i$-th brane, and
$g$ ($h$) is the 5D (4D) metric density.
For later convenience, we define
$\kappa^2 = 8\pi G_N = 1/M_*^3$, where 
$M_*$ is the fundamental 5D Planck mass.

We adopt a conformally flat metric ansatz,
\begin{equation}
ds^2 = g_{MN}dx^Mdx^N
= B(y)(\eta_{\mu\nu}dx^\mu dx^\nu +dy^2)\,.
\label{eq=metric}
\end{equation}
Here, we assume that the 3-brane is flat.

We take the scalar field to reside in the
vacuum manifold, $\Phi^a\Phi^a=\eta^2$. 
The scalar field is of the form,
\begin{equation}
\Phi^a = \eta \left[ \cos\chi (y), \sin\chi (y) \right]\,, 
\end{equation}
where $\chi (y)$ is the phase factor in the field space.
In this case, the nonlinear $\sigma$-model approximation is applied,
and the coupling $\lambda$ appearing 
in the action is treated as a Lagrange multiplier.
In this case, the scalar-field equation is
\begin{equation}
\nabla^A\partial_A\Phi^a  
=-{(\partial^A\Phi^b)(\partial_A\Phi^b) \over \eta^2}\Phi^a
+{\sqrt{-h} \over \sqrt{-g}}
{\partial\sigma_i(\Phi^a) \over \partial\Phi^a}\delta(y_i)
\,,
\end{equation}
where we have included the contribution of the branes.

The energy-momentum tensor of the texture and the branes is given by
\begin{equation}
T^M_N = \partial^M\Phi^a \partial_N\Phi^a
-{1\over 2}\delta^M_N\partial_A\Phi^a\partial^A\Phi^a
-{\sqrt{-h} \over \sqrt{-g}} \sigma_i(\Phi^a)\delta (y_i)
\text{diag} (1,1,1,1,0)
\,.
\end{equation}

The Einstein's equation for the given metric and  energy-momentum 
tensor leads to
\begin{eqnarray}
-G^\mu_\mu &=& {1\over B} \left[ -{3\over 2}{B'' \over B}
+{3\over 4}\left( {B'\over B} \right)^2\right]
= \kappa^2\eta^2 {\chi'^2 \over 2B}
+ {\kappa^2 \over \sqrt{B}} \left[ \sigma_I(\Phi^a)\delta(y)
+ \sigma_{II}(\Phi^a)\delta(y-y_*) \right] \,,\\
-G^y_y &=& {1\over B} \left[ -{3\over 2}\left( {B'\over B} \right)^2\right]
=-\kappa^2\eta^2 {\chi'^2 \over 2B}\,.
\end{eqnarray}
Here, we assumed that there are two branes located at $y=0$ and $y=y_*$,
which will be justified in the next section.

The scalar-field equation leads to
\begin{equation}
\chi'' + {3\over 2}{B'\over B}\chi'
= {2\sqrt{B} \over \eta^2}\left[ 
{\partial\sigma_I \over \partial\chi}\delta(y)
+{\partial\sigma_{II} \over \partial\chi}\delta(y-y_*)
\right]\,.
\label{eq=sfd}
\end{equation}

\section{The solutions}\label{sec=sol}
The solutions to the Einstein and the scalar-field equations 
presented in the previous section are
\begin{eqnarray}
B(y) &=& \left[b_0\chi_0(|y|+y_0)\right]^{2/3}\,,
\label{eq=B}\\
\chi (y) &=& {\text{sign} (y) \over b_0}\ln\left( {|y| \over y_0} +1 \right)\,,
\label{eq=chi}
\end{eqnarray}
where $b_0 = (\sqrt{3}/2) \kappa\eta$. $y_0$ and $\chi_0$
are the integration constants.
For convenience,
we define a new constant, $B_0 =b_0\chi_0$.\footnote{
If we perform a coordinate transformation
to recast the solution  into the
form which contains the warp factor, 
the metric becomes
\begin{equation}
ds^2 = \left[{4\over 3}B_0(|z|+z_0)\right]^{1/2}
\eta_{\mu\nu}dx^\mu dx^\nu  +dz^2\,.\nonumber
\end{equation}
}
\footnote{This type of solutions appear in the 5D model
with a free scalar field. 
(See, for example, Ref.~\cite{Csaki}.)
However, as it will be discussed in the next few paragraphs,
the texture imposes distinctly different boundary conditions
which, for instance, allow the geometry to avoid a naked singularity.
}
We can take $y_0>0$ and $B_0>0$ without changing physics.

\begin{figure}
\begin{center}
\epsfig{file=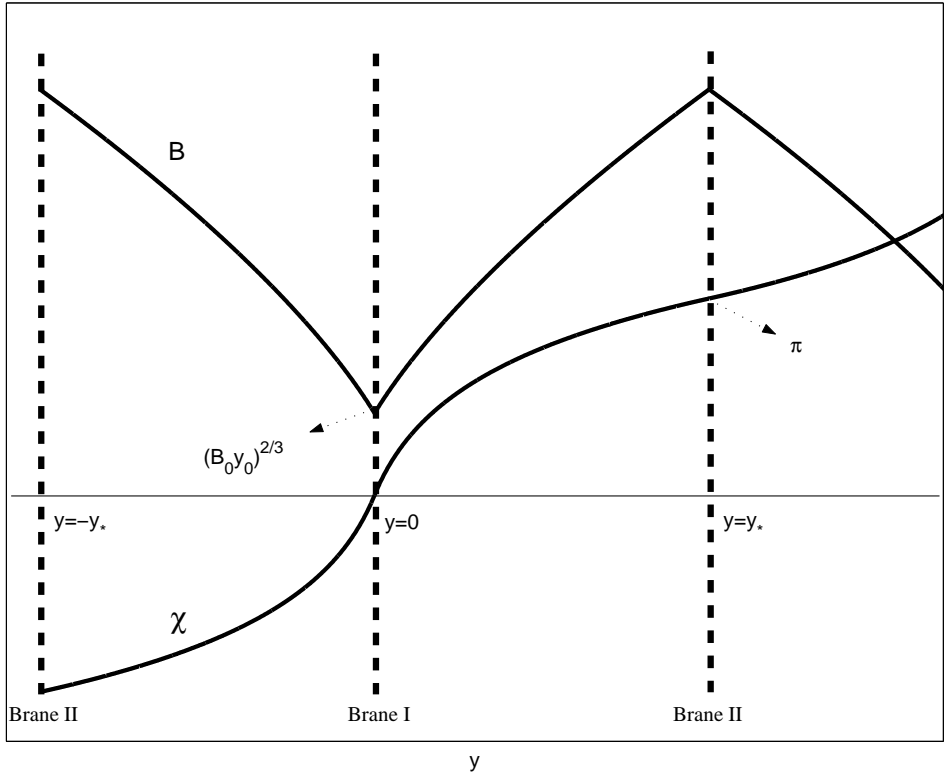,width=5in}
\end{center}
\vspace{0.5in}
\caption{
A schematic picture of the model and solutions.
Two branes are located at $y=0$ and $y=y_*$.
Two points, $y=y_*$ and $y=-y_*$ are identified,
and the extra dimension is compact.
The gravitational field $B$ is lowest on the first brane,
and highest on the second brane.
The scalar field $\chi$ is continuous across the second brane.
}
\label{fig=brane}
\end{figure}

Let us discuss the solutions in detail. 
The solutions are plotted in Fig.\ref{fig=brane}.
In order to fix one integration constant in $\chi (y)$,
we have already set $\chi (y=0) =0$.
The phase factor
$\chi$ varies from $-\pi$ to $\pi$
for the scalar field
to complete one winding in the vacuum manifold.
We set the boundary as
$\chi (y=y_*) =\pi$,
and $\chi (y=-y_*) = -\pi$,
which gives, 
\begin{equation}
y_*=y_0 (e^{b_0\pi}-1) 
\equiv y_0 (e^{3n}-1)\,,
\label{eq=y*}
\end{equation}
where we defined $n=b_0\pi /3 =(\pi /2\sqrt{3})\kappa\eta$.
(We shall refer to $n$ as the symmetry-breaking scale
from now on.)
Since we assume the $S^1/Z_2$ orbifold,
$y=y_*$ and $y=-y_*$ are identified, where
their $\chi$-value corresponds to the same point
in the vacuum manifold.

The gravitational field $B(y)$ is chosen to preserve
$Z_2$ symmetry about $y=0$ and $y=y_*$.
Also, $\chi (y)$ is chosen so that the physical quantities
determined by it should preserve $Z_2$ symmetry.
Then, the observers at $y=0$ and $y=y_*$ see that the extra dimension 
is symmetric.
Such a symmetry can be obtained 
when we place two 3-branes of nonzero tension
at $y=0$ and $y=y_*$.

When we identify $y=y_*$ with $y=-y_*$,
$\chi$ will not be continuous there.
However, this is not a problem
since $\chi = \pi$ and $\chi = -\pi$ represent
the same point in the vacuum manifold.
Therefore, $\chi$ is continuous  on the brane
at $y=y_*$ as seen in Fig.\ref{fig=brane}.
In this sense, this texture of one winding number
can be regarded equivalently as $1/w$ section
of the texture of $w$ winding number.

Let us evaluate the brane tension from the boundary
conditions on the branes.
If we plug the solutions (\ref{eq=B}) and (\ref{eq=chi}) in
the Einstein equations, we obtain the brane tensions at two boundaries.
The tension of the first brane at $y=0$ is given by
\begin{equation}
\sigma_I = -{2 \over \kappa^2(B_0y_0)^{1/3}y_0} <0\,.
\label{eq=sigma1}
\end{equation}
We note that the tension is always negative.

To evaluate the tension of the second brane at $y=y_*$,
we recast the solutions in a slightly different form,
\begin{eqnarray}
B(y) &=& [B_0(-|y-y_*|+y_*+y_0)]^{2/3}\,,\\
\chi (y) &=& -\text{sign} (y-y_*) \left[ {1\over b_0}
\ln \left( {-|y-y_*|+y_* \over y_0} +1 \right) -\pi 
\right] +\pi\,.
\end{eqnarray}
Note that the solutions are unchanged in the bulk.
From the Einstein equations, the tension of the second brane
is given by
\begin{equation}
\sigma_{II} ={2 \over \kappa^2[B_0(y_0+y_*)]^{1/3}(y_0+y_*)}
={2\over \kappa^2(B_0y_0)^{1/3}y_0e^{4n}} >0\,,
\label{eq=sigma2}
\end{equation}
which is always positive.
The ratio of the tensions is $|\sigma_I /\sigma_{II}| =e^{4n}$.
$\sigma_I$  always has the larger magnitude.

In addition, from the scalar-field equation~(\ref{eq=sfd}),
we have
\begin{equation}
{\partial\sigma_I \over \partial\chi} =
{\partial\sigma_{II} \over \partial\chi} = 0\,.
\end{equation}

We find that the tension is negative on the first brane where
the gravitational field $B(y)$ is lowest, and is positive
on the second brane where $B$ is highest.
This is similar to the Randall-Sundrum  model (RSI)~\cite{RSI}.

Now, let us evaluate the proper distance between two branes.
It is given by
\begin{equation}
l = \int_0^{y_*} \sqrt{B(y)}dy
={3\over 4} (B_0y_0)^{1/3}y_0 (e^{4n}-1)
={3\over 2\kappa^2}\left( {1\over \sigma_I} 
+ {1\over \sigma_{II}} \right)\,.
\label{eq=l}
\end{equation}
Remarkably, the proper distance is determined 
by the sum of the inverse tensions of two branes.
Thus, the three physical quantities, $\sigma_I$, $\sigma_{II}$, and $l$, are
related to one another.
Once one of the tensions is determined, the other tension
is automatically determined for a given symmetry-breaking scale ($n$),
and the proper distance between the branes is set.

With the given solution for the gravitational field,
let us discuss the particle hierarchy of the model.
We follow the same reasoning that was used in Ref.~\cite{RSI}.
The induced metric on the brane located at $y=y_i$ is 
\begin{equation}
g^i_{\mu\nu} = g_{MN}(x^\mu ,y=y_i) 
= [B_0(y_i+y_0)]^{2/3}\eta_{\mu\nu}\,,
\end{equation}
and the metric on the second brane is
\begin{equation}
g^{II}_{\mu\nu} = g_{MN}(x^\mu ,y=y_*) 
= (B_0y_0)^{2/3}e^{2n} \eta_{\mu\nu}\,.
\end{equation}
Let us consider Higgs-like particles on the brane at $y_i$.
The action containing such particles is written as
\begin{eqnarray}
{\cal S}_i \supset &-& \int d^4x \sqrt{-g^i} [g_i^{\mu\nu} D_\mu H^\dag D_\nu H
+\lambda (|H|^2-v_i^2)^2 ] \\
=&-& \int d^4x\sqrt{-g^{II}} \left[ \left( {y_i\over y_0}+1\right)^{2/3}e^{-2n}
g_{II}^{\mu\nu} D_\mu H^\dag D_\nu H 
+\left({y_i\over y_0}+1\right)^{4/3}e^{-4n} \lambda (|H|^2 -v_i^2)^2 \right]\,.
\end{eqnarray}
If we rescale the Higgs field by 
$\overline{H} = (y_i/y_0 +1)^{1/3}e^{-n}H$, the effective action
which describes the particles on the brane at $y_i$ is given by
\begin{equation}
{\cal S}_{eff} = -\int d^4x \sqrt{-g^{II}} [ g_{II}^{\mu\nu}
D_\mu \overline{H}^\dag D_\nu \overline{H} +\lambda (|\overline{H}|^2-v_{eff}^2)^2 ]\,.
\end{equation}
The particles on the brane at $y_i$ are effectively
described by the metric of the second brane $g^{II}$, and the
effective symmetry-breaking scale becomes
\begin{equation}
v_{eff} = \left( {y_i\over y_0} + 1 \right)^{1/3}e^{-n}v_i\,.
\label{eq=veff}
\end{equation}
The effective-particle scale is largest on the second brane,
and flows along the extra dimension by a power law.
The particle hierarchy between the two branes at $y=0$ and $y=y_*$
is completely determined by the symmetry-breaking scale of the texture.
We assume that Planck-scale particles are confined to the second brane.
If we wish the particle scale on the first brane to be of TeV,
it requires $n\sim 35$ ($\kappa\eta \sim 38$).
Therefore,
the model requires only that the symmetry-breaking scale 
should be slightly supermassive
to the fundamental 5D Planck mass scale.

From Eq.~(\ref{eq=veff}) we observe that the desired particle hierarchy
is set up only if the location of the brane $y_i$ is very close
to the first brane at $y=0$. 
This is because the gravitational field is a power-law function of $y$, 
rather than an exponentional one.
From now on, we assume that our world is located on the first, TeV, brane.

\section{Localization of gravity}\label{sec=zero}
In this section, we investigate the localization of gravitons.
Let us consider a small perturbation $h_{\mu\nu}$ to the background
metric,
\begin{equation}
ds^2 = B(y)\left\{ \left[ \eta_{\mu\nu} + {h_{\mu\nu}\over B(y)}
\right]dx^\mu dx^\nu +dy^2\right\}\,.
\label{eq=metrich}
\end{equation}
We assume that the only nonzero components are $h_{\mu\nu}$
($h_{\mu y} = h_{yy} =0$),
and apply a transverse-traceless gauge on $h_{MN}$, 
$h_{MN|}{}^N=0$ and $h^M_M =0$, where the vertical bar
denotes the covariant derivative with respect to the background metric.
The Einstein's equation for $h_{MN}$ reads
\begin{equation}
h_{MN|A}{}^A +2R^{(B)}_{MANB}h^{AB}
-2R^{(B)}_{A(M}h^A_{N)}
+{2(6-N)\over N-2}\kappa^2\sigma_i\delta (y_i)h_{MN}=0\,.
\label{eq=h}
\end{equation}
Here, the superscript $(B)$ represents that the quantity is 
evaluated in the unperturbed background metric,
and $N=5$.
We search the solution of a form, $h_{\mu\nu} =h(y)\hat{e}_{\mu\nu}
e^{ip_\mu x^\mu}$.
In order to cast Eq.~(\ref{eq=h}) into a nonrelativistic
Schr\"odinger-type equation, we define a new function by
$\hat{h}(y) = B^{-1/4}h(y)$, then the equation leads to
\begin{equation}
\left[ -{1\over 2}{d^2 \over dy^2} + V(y) \right]\hat{h}(y)
={m^2 \over 2}\hat{h}(y)\,,
\label{eq=hhat}
\end{equation}
where the 4D graviton mass is determined by 
$\Box^{(4)}h =-p_\mu p^\mu h =m^2h$, and the potential is 
\begin{equation}
V(y) = -{1\over 8(|y|+y_0)^2}-\Sigma_I\delta (y)\,,
\label{eq=V1}
\end{equation}
on the first brane, and
\begin{equation}
V(y) = -{1\over 8(|y-y_*|-y_*-y_0)^2}+\Sigma_{II}\delta (y-y_*)\,,
\label{eq=V2}
\end{equation}
on the second brane.
Note that the $V$'s are the same in the bulk.
The coefficients in the front of the delta functions are
\begin{eqnarray}
\Sigma_I &=& {1\over 6y_0} + {1\over 3}\kappa^2(B_0y_0)^{2/3}\sigma_I
= {1 \over 6y_0}\left[ 1- 4(B_0y_0)^{1/3} \right]
\,,
\label{eq=Sigma1}\\
\Sigma_{II} &=& {1\over 6(y_0+y_*)} 
- {1\over 3}\kappa^2[B_0(y_0+y_*)]^{2/3}\sigma_{II}
= {1 \over 6y_0e^{3n}}\left[ 1- 4(B_0y_0)^{1/3}e^n \right]
\,.
\label{eq=Sigma2}
\end{eqnarray}

For the zero mode, $m=0$, the solution to the equation~(\ref{eq=hhat}) is
\begin{equation}
\hat{h}^I_0(y) = \sqrt{|y|+y_0} [a_1 +a_2 \ln (|y|+y_0)]\,.
\label{eq=h00}
\end{equation}
Here, the superscript $I$ denotes that the solution is valid
in the bulk and on the first brane. 
If we consider the original form of the perturbation to the 
background metric, this solution gives
\begin{equation}
ds^2 = B(y) \left\{ [ \eta_{\mu\nu}+ [ b_1 +b_2\chi (y)] \hat{e}_{\mu\nu}e^{ip_\mu x^\mu}]
dx^\mu dx^\nu +dy^2 \right\}\,.
\end{equation}
The first term of the zero-mode wave function
turns out to be a constant contribution to the perturbation.
This is analogous to that the  amplitude of
tensor perturbations stays constant
on super-horizon scales 
in the standard four dimensional theory.
This type of the solution is the only zero mode 
that arises in the other models.
The second term of the wave function
is the contribution of the scalar field $\chi (y)$
to the perturbation.

Note that neither of these two terms solely satisfy the
boundary conditions on the branes. 
The zero mode must be  a combination of these two terms.
This is not true in the other models.
Their boundary conditions usually require
only one term in the zero-mode wave function.

\begin{figure}
\begin{center}
\epsfig{file=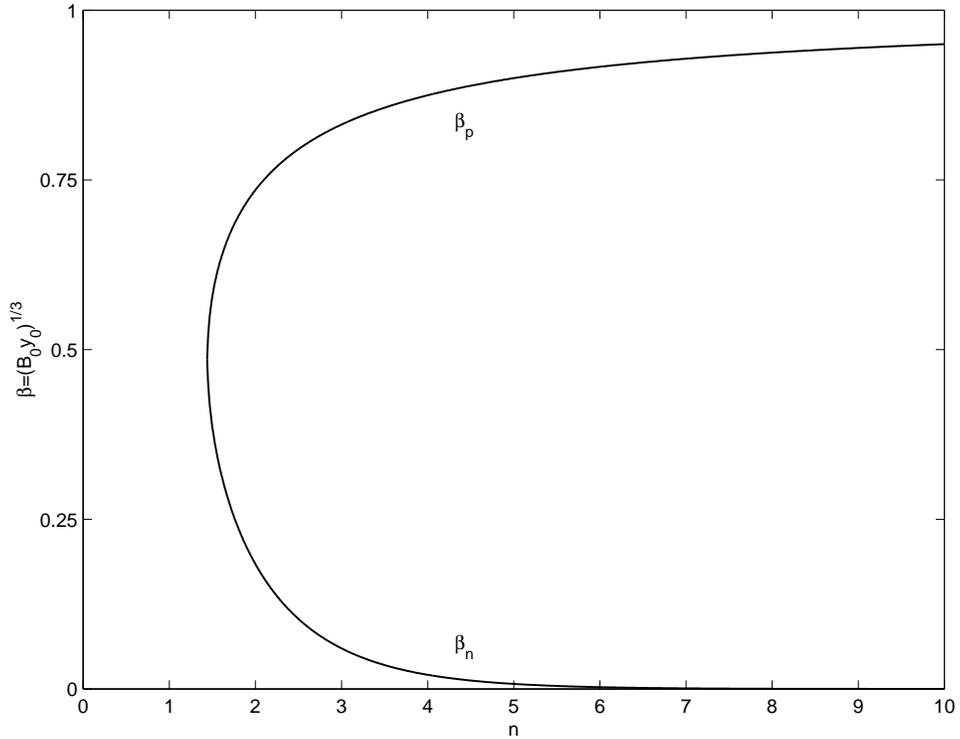,width=5in}
\end{center}
\vspace{0.5in}
\caption{
Solutions for $\beta =(B_0y_0)^{1/3}$ as a function of $n$.
There are two solutions for a given $n\ge n_c \simeq 1.44$.
We call the upper branch $\beta_p$ and the lower $\beta_n$.
The solutions are always in the range,  $0<\beta <1$.
}
\label{fig=beta}
\end{figure}

Now, let us determine the coefficients of the solution
considering boundary conditions on two branes.
In order to consider the boundary condition on the second brane
at $y=y_*$, we need to recast the solution in
\begin{equation}
\hat{h}^{II}_0(y) = \sqrt{-|y-y_*|+y_*+y_0} [a_1 +a_2 
\ln (-|y-y_*|+y_*+y_0)]\,.
\label{eq=h0*}
\end{equation}
If we plug solutions (\ref{eq=h00}) and (\ref{eq=h0*})  
into Eq.~(\ref{eq=hhat}), 
we get two conditions at the boundaries, $y=0$ and $y=y_*$, 
\begin{equation}
-{a_1\over a_2} = {2 \over 1+2\Sigma_Iy_0}+\ln y_0
= {2 \over 1+2\Sigma_{II}y_0 e^{3n}}+\ln (y_0e^{3n})\,.
\label{eq=a1a2}
\end{equation}
These conditions eliminate one of the two constants 
($a_1$ or $a_2$) in the wave function, and
determine $(B_0y_0)^{1/3}$.
Using the expressions for $\Sigma_I$ and $\Sigma_{II}$,
in Eqs.~(\ref{eq=Sigma1}) and (\ref{eq=Sigma2}), and by
defining $\beta\equiv (B_0y_0)^{1/3}$, 
the above condition reduces to
\begin{equation}
e^n\beta^2 - \left[ 1+e^n+ {1\over 2n}(1-e^n) \right]\beta +1 =0\,.
\label{eq=beta}
\end{equation}
Therefore, $\beta$ is a function of $n$ only, and is determined
by solving this polynomial equation. 
The solution is plotted in Fig.\ref{fig=beta}.
The solution exists above some critical value $n \geq n_c\simeq 1.44$.
There are two solutions of $\beta$ for a given $n$, and the
solutions are in the range, $0<\beta <1$. 
We call the upper branch of the solution $\beta_p$, and the lower $\beta_n$.

Applying the normalization condition, 
$\int^{y_*}_0 |\hat{h}_0(y)|^2 dy = e_l$,  
determines the remaining constant of the wave function, 
where $e_l$ is the unit length.
The zero-mode wave function is then
\begin{equation}
\hat{h}_0(y) =  [2y_0^2(\alpha_1 e^{6n} +\alpha_2)/e_l]^{-1/2}
\sqrt{|y|+y_0} \left[-{2\over s_1} + \ln\left( {|y| \over y_0}
+1\right)\right]\,,
\label{eq=h0complete}
\end{equation}
where
\begin{eqnarray}
s_1 &=& 1+2\Sigma_Iy_0
={4\over 3}\left[ 1- (B_0y_0)^{1/3}\right]
\,,\label{eq=s1}\\
s_2 &=& 1+2\Sigma_{II}y_0e^{3n}
={4\over 3}\left[ 1- (B_0y_0)^{1/3}e^n\right]
\,,\label{eq=s2}\\
\alpha_1 &=& {2\over s_1^2} +{1\over s_1} +{1\over 4}
+{9n^2\over 2}-{6n\over s_1}-{3n\over 2}\,,\\
\alpha_2 &=& -\left( {2\over s_1^2} +{1\over s_1} +{1\over 4} \right)\,.
\end{eqnarray}
All of these constants are the function of  $n$ only, via $\beta(n)$.
$s_2$ is defined here for  future convenience.

\begin{figure}
\begin{center}
\epsfig{file=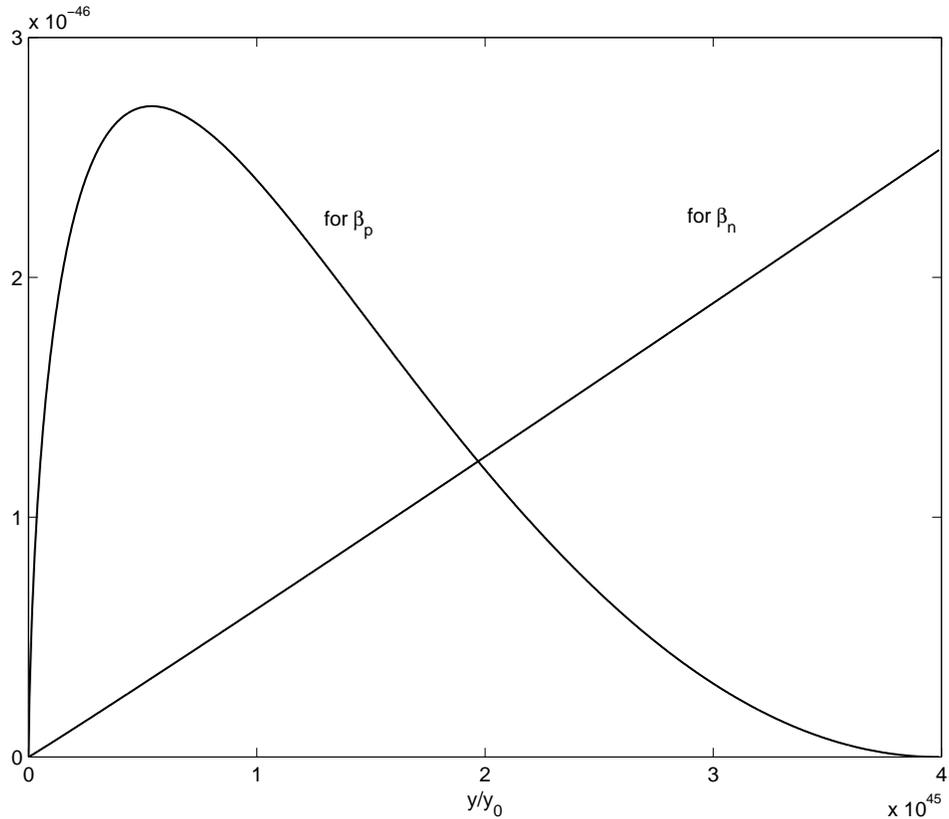,width=5in}
\end{center}
\vspace{0.5in}
\caption{
A plot of the squared amplitude of the zero-mode wave function,
$y_0|\hat{h}_0(y)|^2$, for $n=35$.
For the $\beta_p$-solution, the amplitude is largest
somewhere between the two branes. 
For the $\beta_n$-solution, the amplitude is largest
on the second brane.
For the $\beta_p$-solution,
the amplitude on the second brane is much larger than that
of the first brane ($y=0$) even though we can hardly see it in the plot.
There is a location ($y \neq 0, y_*$)
where the amplitude vanishes for each solution.
For $\beta_p$-solution, it is close to the second brane, and
for $\beta_n$, close to the first brane.
}
\label{fig=h0}
\end{figure}

The ratio between the amplitudes on the first and second branes is
\begin{equation}
{|\hat{h}_0(y_*)|^2  \over |\hat{h}_0(0)|^2}
=e^{3n}\left( 1-{3\over 2}s_1n \right)^2\,.
\end{equation}
For $n=35$, the amplitude of the zero mode is much larger on the
second brane than the first brane.
The amplitude of the wave function is plotted in  Fig.\ref{fig=h0}. 
For the $\beta_n$ solution, the amplitude is largest on the second brane,
while for the $\beta_p$ solution, it is largest somewhere between the
first and the second brane.
There exists a point where the amplitude vanishes
for both solutions.

With the zero-mode wave function,
the Newtonian gravitational potential between the
test mass $M_1$ and $M_2$ separated by  the distance $r$
on the brane, is given by
\begin{equation}
V_{Newt}(r) = { G_N |\hat{h}_0(y)|^2 \over \int |\hat{h}_0(y)|^2 dy}
{M_1M_2 \over r}
\end{equation}
The 4D gravitational constant on the first brane is given by
\begin{equation}
G_4(y=0) = G_N {|\hat{h}_0(0)|^2 \over e_l} = 
G_N {2 \over y_0s_1^2 (\alpha_1e^{6n} + \alpha_2)}\,.
\end{equation}
Since $G_4(0)=1/M_{Pl}^2$ and $G_N=1/M_*^3$,
$y_0$ is expressed by the fundamental mass scale
\begin{equation}
y_0 = {1\over M_*}\left( M_{Pl} \over M_* \right)^2
{2 \over s_1^2(\alpha_1e^{6n}+\alpha_2)}\,.
\label{eq=y0}
\end{equation}
If we require that the fundamental 5D Planck mass $M_*$ should be of TeV
scale, the above formula determines the value of $y_0$.

Therefore, all the constants in the model have been determined. 
Requiring the particle hierarchy between two branes,
the symmetry-breaking scale of the texture was determined ($n\sim 35$).
The boundary condition of the zero-mode graviton fixes $\beta =(B_0y_0)^{1/3}$
for a given $n$.
Finally, $y_0$ is determined by the requirement of the hierarchy
between the 4D and the 5D Planck mass.

Since we have determined all the constants, let us estimate some physical
quantities determined by those constants. The quantities are shown in
Table~\ref{tab=scale} for $n=35$.
Since there are two values of $\beta$ for a given $n$, two values for
each quantity have been presented. 
$\beta$ and $y_0$ are evaluated by Eq.~(\ref{eq=beta}) and Eq.~(\ref{eq=y0}).
The large difference between $\sigma_I$ and $\sigma_{II}$ is already
implied in their formulae, Eqs.~(\ref{eq=sigma1}) and (\ref{eq=sigma2}).
Since the proper distance $l$ between the two branes is determined
by the sum of the inverse tensions, its numerical scale is mainly
determined by the inverse of the smaller tension, $\sigma_{II}$.

For $\beta_p$, the proper distance $l$ is reasonably large.
This separation will not be disturbed by  quantum fluctuations
of the particles on the TeV brane. 
On the other hand, for $\beta_n$,  
the separation is far smaller than the particle-length scales
on the brane. 
This length scale will be easily disturbed.
Therefore, from now on, we shall consider  only the $\beta_p$-solution.

\section{Massive Kaluza-Klein gravitons}\label{sec=KK}
In this section, we study  massive Kaluza-Klein (KK) gravitons.
Since our model is a compactified one, there exists
a discrete spectrum of massive modes.
The solution of the mass eigenvalue equation~(\ref{eq=hhat}) is given by
\begin{equation}
\hat{h}^I_m(y) = \sqrt{|y|+y_0} \left\{ c_1 J_0[m(|y|+y_0)] 
+ c_2 Y_0[m(|y|+y_0)]\right\}\,,
\label{eq=hm0}
\end{equation}
where $J$ ($Y$) is the Bessel function of the first (second) kind.
In the limit of small argument of the Bessel function, 
$m(|y|+y_0) \ll 1$, 
the first term reduces to the first term of the zero-mode wave function, 
and the second does to the second.
Therefore, the zero-mode wave function is a limit of the
massive-mode wave function when $m\to 0$.
This fact will help us later in analyzing the correction to the Newtonian 
gravitational potential from very light modes.

Viewed from the second brane, the solution is recast in the form,
\begin{equation}
\hat{h}^{II}_m(y) = \sqrt{-|y-y_*|+y_*+y_0} 
\left\{ c_1 J_0[m(-|y-y_*|+y_*+y_0)] 
+ c_2 Y_0[m(-|y-y_*|+y_*+y_0)]\right\}\,.
\label{eq=hm*}
\end{equation}
Similarly to the zero-mode case,
on the two branes, 
these solutions (\ref{eq=hm0}) and (\ref{eq=hm*})
provide two conditions from Einstein equations. 
The ratio of the coefficients is determined by
\begin{equation}
-{c_1 \over c_2} =
{s_1Y_0(my_0) -2my_0Y_1(my_0) \over s_1J_0(my_0) -2my_0J_1(my_0)}
={s_2Y_0(my_0e^{3n}) -2my_0e^{3n}Y_1(my_0e^{3n}) 
\over s_2J_0(my_0e^{3n}) -2my_0e^{3n}J_1(my_0e^{3n})}\,,
\label{eq=c1c2}
\end{equation}
where $s_1$ and $s_2$ were defined in Eqs.~(\ref{eq=s1}) and 
(\ref{eq=s2}).
Then, the wave function reads
\begin{equation}
\hat{h}^I_m(y) =c_2 \sqrt{|y|+y_0} \left\{ 
-{s_1Y_0(my_0) -2my_0Y_1(my_0) \over s_1J_0(my_0) -2my_0J_1(my_0)}
J_0[m(|y|+y_0)] 
+  Y_0[m(|y|+y_0)]\right\}\,,
\label{eq=hm02}
\end{equation}
where $c_2$ is determined by the normalization,
\begin{eqnarray}
c_2^2/e_l &=& \left\{ 
y_0^2e^{6n}(s_2^2+4m^2y_0^2e^{6n})
{[J_0(my_0e^{3n})Y_1(my_0e^{3n})-J_1(my_0e^{3n})Y_0(my_0e^{3n})]^2
\over
[s_2J_0(my_0e^{3n})-2my_0e^{3n}J_1(my_0e^{3n})]^2}\right. \nonumber\\
{} &-& \left.
y_0^2(s_1^2+4m^2y_0^2)
{[J_0(my_0)Y_1(my_0)-J_1(my_0)Y_0(my_0)]^2
\over
[s_1J_0(my_0)-2my_0J_1(my_0)]^2} 
\right\}^{-1}\,.
\end{eqnarray}
The wave function $\hat{h}_m^{II}(y)$ is written in a similar way.

\begin{figure}
\begin{center}
\epsfig{file=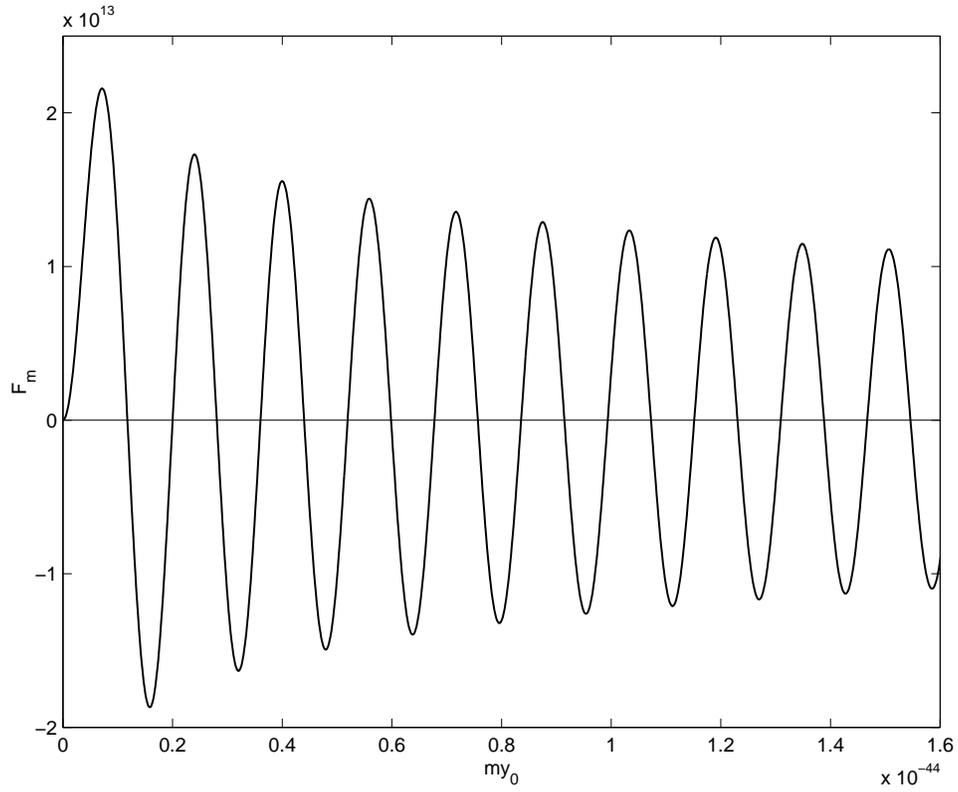,width=5in}
\end{center}
\vspace{0.5in}
\caption{
Plot of the mass-spectrum function $F_m$ for $n=35$.
The zeroes of the function corresponds to the mass eigenvalues
of KK gravitons.
}
\label{fig=Fm}
\end{figure}

In addition to determining the ratio $c_1/c_2$, 
Eq.~(\ref{eq=c1c2}) determines the Kaluza-Klein mass spectrum.
Unfortunately, it seems impossible to get an analytic description
for $m$ from the above equation.
In order to view the mass spectrum, 
we construct a mass-spectrum function from Eq.~(\ref{eq=c1c2}),
\begin{eqnarray}
F_m(my_0) &=& [s_1Y_0(my_0) -2my_0Y_1(my_0)]
[s_2J_0(my_0e^{3n}) -2my_0e^{3n}J_1(my_0e^{3n})]\nonumber\\
&-& [s_1J_0(my_0) -2my_0J_1(my_0)]
[s_2Y_0(my_0e^{3n}) -2my_0e^{3n}Y_1(my_0e^{3n})] \,.
\label{eq=Fm}
\end{eqnarray}
The zeroes of this function give the mass eigenvalues.
It is plotted in Fig.\ref{fig=Fm} for $n=35$ and for $\beta_p$.
A few of the light modes are shown in the plot.
As usual in the compactified model,
the mass gap is about the inverse of the extra dimension size,
$\Delta m \sim  1/y_* \approx 1/y_0e^{3n}$.

Now, let us discuss the contribution of the light KK modes
to the Newtonian gravitational potential.
The correction from the KK modes on the first brane is given by
\begin{equation}
\Delta V_{Newt}(r) = {M_1M_2 \over r} \sum_j
{G_N |\hat{h}_m(0)|^2 \over e_l}
e^{-m_jr}\,,
\label{eq=VKK}
\end{equation}
where $e_l$ comes from the normalization of the wave function.
As it was mentioned in the beginning of this section,
in the limit of $m(|y|+y_0) \ll 1$, the wave function $\hat{h}_m(y)$
reduces to the zero-mode wave function $\hat{h}_0(y)$.
If we consider the mass gap, $\Delta m(y_*+y_0) \sim 1$,
the above limit is achieved only when $y$ is small:
At the location of the second brane, the above limit cannot
be achieved.
Hence, on the first brane,  for very light KK modes,
the wave function is very close to that of the zero mode.
The KK coupling constant is then approximately
the zero-mode coupling which is the 4D Planck scale,
$G_N |\hat{h}_m(0)|^2/e_l \approx G_N |\hat{h}_0(0)|^2/e_l$.
The light KK modes couple as weakly as the zero mode does on the first brane.

Even if the coupling is weak, the correction should be suppressed
by the exponential Yukawa factor.
Otherwise,  its contribution 
to the Newtonian potential will not be negligible.
The exponential factor is negligible when $m_jr \gg 1$, therefore,
\begin{equation}
r \gg  {1\over m_j} \sim {y_0e^{3n} \over j} 
\sim {10^{-11}\over j} \text{TeV}^{-1}
\approx {10^{-27} \over j} \text{mm} \,.
\end{equation}
This range is much smaller than the current experimental resolution
which is of order 1mm.
Therefore, the effect of the light KK modes to the Newtonian potential
is negligible in the regime where gravity is measured currently.

To study the very massive KK modes,
first, let us evaluate the mass spectrum more precisely.
For the large-mass limit, $my_0 \gg 1$, 
the spectrum function $F_m$ in Eq.~(\ref{eq=Fm}) reduces to
\begin{equation}
F_m \approx {\pi my_0 \over 2} \left\{ 
(s_1s_2 + 4m^2y_0^2e^{3n})\sin[my_0(1-e^{3n})]
-2my_0(s_1e^{3n}-s_2)\cos[my_0(1-e^{3n})] \right\} \,,
\end{equation}
where we have used the asymptotic formulae of the Bessel functions 
for the large argument, 
$J_n(x) \approx \sqrt{2/\pi x} \cos(x-n\pi /2 -\pi /4)$, and
$Y_n(x) \approx \sqrt{2/\pi x} \sin(x-n\pi /2 -\pi /4)$.
In order to obtain the mass spectrum, we require $F_m=0$
which reduces to
\begin{equation}
\tan[my_0(e^{3n}-1)] = -{2my_0(s_1e^{3n}-s_2) \over
s_1s_2 + 4m^2y_0^2e^{3n} }\,.
\end{equation}
Then, for large $m$, the mass spectrum is approximately given by
\begin{equation}
m_j \approx {j\pi \over y_0(e^{3n}-1)} = {j\pi \over y_*}\,.
\end{equation}
This spectrum provides approximately the same mass gap that
we guessed for the light modes.

With this mass spectrum, the sum in the correction to the Newtonian
potential~(\ref{eq=VKK}) converges.
Similarly to the light modes, the Yukawa factor suppresses 
the correction efficiently 
when $r \gg (10^{-27} /j\pi )\text{mm}$. 
Therefore, the contribution of the very massive KK modes to the
Newtonian potential is also very small.

To estimate the KK coupling constant of the very massive modes,
we obtain an approximate description of $\hat{h}_m(y)$ from
Eq.~(\ref{eq=hm02}).
Then, we find
\begin{equation}
G_N{|\hat{h}_m(0)|^2 \over e_l} \sim {G_N \over y_*} 
\approx {10^{11} \over M_*^2}\,.
\end{equation}
This is almost an MeV-scale coupling which is very large.
However, as it was shown earlier,
the potential is strongly suppressed by the Yukawa factor, 
so the total correction is very small
within the experimental resolution.

Before we close this section, let us investigate if there are any
tachyonic-graviton modes.
Usually the solutions of a Sturm-Liouville type equation
possess $n$-nodes for a given eigenvalue $E_n$. 
There exists the lowest eigenvalue $E_0$ 
of which the corresponding eigenfunction does not have any zero. 
The zero-mode wave function~(\ref{eq=h0complete}) of our model
possesses one zero regardless of the value of the symmetry-breaking
scale $n$.
Therefore, we suspect there exists one tachyonic mode, $m^2<0$,
which is the lowest mode of the solutions.

\begin{figure}
\begin{center}
\epsfig{file=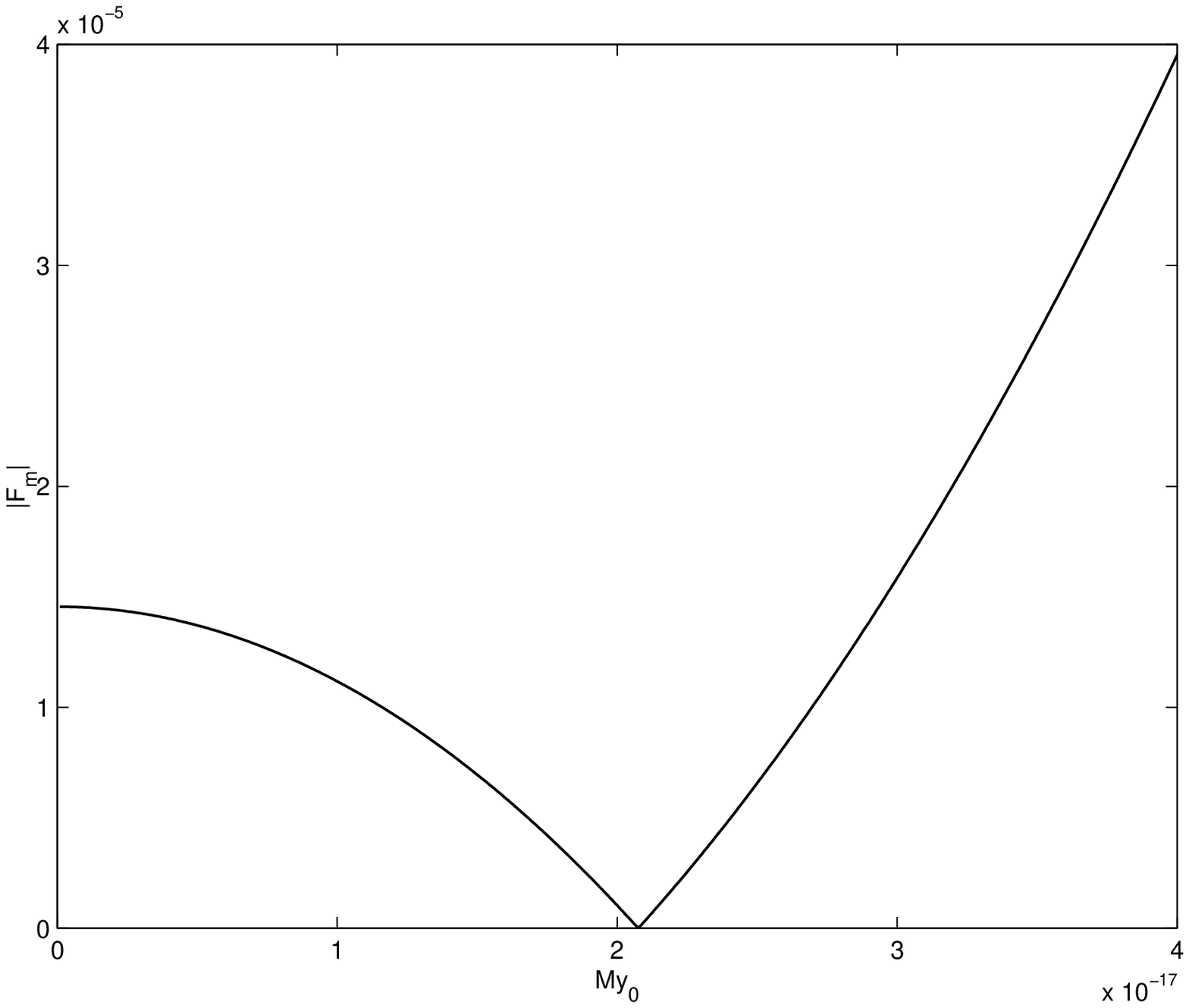,width=5in}
\end{center}
\vspace{0.5in}
\caption{
Plot of the mass-spectrum function $|F_m|$ for tachyonic modes.
The numerical result is shown for $n=10$ instead of for $n=35$
because of the numerical difficulties related with the large numbers
involved for $n=35$.
There is one zero of $|F_m|$ which represents the tachyonic mode.
}
\label{fig=Ft}
\end{figure}

In order to see if there exists a tachyonic mode,
we may use exactly the same analysis as we made for the regular modes.
For tachyonic modes, the mass-spectrum function $F_m$ in 
Eq.~(\ref{eq=Fm}) becomes complex after we replace $m=iM$,
where $M$ is real.
The zero of $|F_m|$ corresponds to the tachyonic mode if there is any.
We perform numerical calculations to search the zero of $|F_m|$.
There result shows that there is one zero of $|F_m|$. 
(See Fig.\ref{fig=Ft}.)

Since there exists a tachyonic mode in the gravitational perturbations,
the brane becomes unstable.
Even if we found the static solutions of the model, 
the stability is disturbed by the growing perturbation
due to the tachyonic mode.
However, it is too early to conclude that the model is completely
unstable.
We have considered only the tensorial part of gravitational perturbations.
In order to investigate the stability in full considerations, 
we need to include the perturbations of the scalar field as well as
the radion~\cite{Charmousis,Garriga}.
If we take those perturbations into an account,
the story of the stability may change.

\section{Conclusions}\label{sec=con}
We investigated a brane model in five dimensions
provided by an O(2) texture formed in one extra dimension.
It has  two branes of nonzero tension, 
and the extra dimension is compact.
One brane has a positive tension, and the other has a negative
tension.
In order to satisfy the boundary conditions on the two branes,
the symmetry-breaking scale of the texture
needs to be slightly super-Planckian.
The gravitational field (or warp factor) is a power-law 
function of the extra coordinate, $y$. (The warp factor
has a power $1/2$ in $y$.)
The warp factor is highest on the positive-tension brane,
and lowest on the negative-tension brane.

Even though the warp factor is not an exponential function of $y$,
the particle hierarchy between two branes is well set up
by an exponential factor
produced solely by the symmetry-breaking scale
of the texture.
In order to achieve a proper particle hierarchy,
the symmetry-breaking scale needs to be $\eta \approx 38/\sqrt{8\pi G_N}$.
Planck-scale particles are then confined to the positive-tension brane, 
and TeV-scale particles are confined to the
negative-tension brane.

The magnitude of the tension is larger  on the negative-tension brane
by an exponential factor of the symmetry-breaking scale.
The proper separation between two branes is completely
determined by the two tensions. 
For the value of $\eta$ given above, it is order of $10^4\text{TeV}^{-1}$.
This separation will be stable against the quantum fluctuations
of the particles on the branes.

Four dimensional gravity is recovered on the TeV brane.
As in the RSI model,  
the amplitude of the zero-mode graviton wave function 
is suppressed on the TeV brane compared with
that on the Planck brane.
In fact, this is how we have a very weak coupling 
of four dimensional gravity in our world. 
However, this suppression of the tensor-mode gravity makes
the scalar-mode gravity more significant, which is not damped.
In this case, 4D gravity deviates from the standard
Einstein gravity to Brans-Dicke gravity~\cite{Garriga}.

Since the extra dimension is compact,
there exists a discrete spectrum of massive Kaluza-Klein modes.
On the TeV brane,
light-mode gravitons couple as weakly as the zero mode does. 
Their contribution to the Newtonian potential 
is suppressed by the Yukawa factor.
The heavy-mode gravitons couple very strongly, at an MeV scale, but
they are again suppressed in the regime where  gravity
is currently measured.
Therefore, the observer on the TeV brane essentially sees gravity
as four dimensional.

Other than the zero mode and the massive Kaluza-Klein modes,
there also exists a tachyonic mode.
Due to this mode the brane will be unstable.
However, to complete discussing the stability of the model,
we need to include the perturbations of the scalar field and the radion.
We hope we get back to this in the future.

\acknowledgements
I am grateful to Katherine Benson for encouraging me working on 
a texture model.
I also thank Christos Charmousis, Stephen Davis, 
Jihad Mourad, and Takahiro Tanaka 
for helpful discussions.

\clearpage

\begin{table}
\begin{center}
\begin{tabular}{ccccc}
$\beta=(B_0y_0)^{1/3}$ & $y_0$                   & $\sigma_I$           & $\sigma_{II}$       & $l$ 
\\ \hline
$\beta_p  =0.986$      & $1.39\times 10^{-57}\text{TeV}^{-1}$ & $-1.46\times 10^{57}\text{TeV}^4$ 
& $2.31\times 10^{-4}\text{TeV}^4$ & $6.48\times 10^3 \text{TeV}^{-1}$
\\ \hline
$\beta_n  =6.46\times 10^{-16}$ & $1.33\times 10^{-65}\text{TeV}^{-1}$ & $-2.35\times10^{80}\text{TeV}^4$ 
& $3.71\times 10^{19}\text{TeV}^4$ & $4.04\times 10^{-20} \text{TeV}^{-1}$
\\ 
\end{tabular}
\end{center}
\caption
{Physical quantities of the model for $n=35$ ($\kappa\eta = 38 $).
There are two values for the given $n$.
}
\label{tab=scale}
\end{table}

\clearpage

\clearpage

\clearpage

\clearpage

\clearpage


\begin{references}
\bibitem{ADD}
N. Arkani-Hamed, S. Dimopoulos and G. Dvali,
Phys. Lett. B {\bf 429}, 263 (1998);
Phys. Rev. D {\bf 59}, 086004 (1999);
I. Antoniadis, N. Arkani-Hamed, S. Dimopoulos, and G. Dvali,
Phys. Lett. B {\bf 436}, 257 (1998).
\bibitem{RSI}
L. Randall and R. Sundrum,
Phys. Rev. Lett. {\bf 83}, 3370 (1999).
\bibitem{RSII}
L. Randall and R. Sundrum,
Phys. Rev. Lett. {\bf 83}, 4690 (1999).
\bibitem{RSIII}
J. Lykken and L. Randall,
JHEP 0006:014, (2000). 
\bibitem{Defects}
V. Rubakov and M. Shaposhnikov,
Phys. Lett. B {\bf 125}, 139 (1983);
{\bf 125}, 136 (1983);
G. Dvali and M. Shifman,
Phys. Lett. B {\bf 396}, 64 (1997);
A. Cohen and D. Kaplan,
Phys. Lett. B {\bf 470}, 52 (1999);
R. Gregory,
Phys. Rev. Lett. {\bf 84}, 2564 (2000);
I. Olasagasti and A. Vilenkin,
Phys. Rev. D {\bf 62}, 044014 (2000);
T. Gherghetta and M. Shaposhnikov,
Phys. Rev. Lett. {\bf 85}, 240 (2000);
T. Gherghetta, E. Roessl and M. Shaposhnikov,
Phys. Lett. B {\bf 491}, 353 (2000);
K. Benson and I. Cho,
Phys. Rev. D {\bf 64}, 065026 (2001);
E. Roessl and M. Shaposhnikov,
hep-th/0205320.
\bibitem{Texture}
For a review of four dimensional textures, see, for example,
R. L. Davis,
Phys. Rev. D {\bf 35}, 3705 (1987);
N. Turok,
Phys. Rev. Lett. {\bf 63}, 2625 (1989);
N. Turok and D. Spergel, 
{\it ibid.} {\bf 64}, 2736 (1990);
M. Barriola and T. Vachaspati,
Phys. Rev. D {\bf 43}, 1056 (1991);
{\it ibid.} {\bf 43}, 2726 (1991);
D. N\"otzold,
Phys. Rev. D {\bf 43}, R961 (1991).
\bibitem{Csaki}
C. Cs\'aki, J. Erlich, C. Grojean, and T. Hollowood,
Nucl. Phys. {\bf B584}, 359 (2000);
S. Kachru, M. Schulz and E. Silverstein,
Phys. Rev. D {\bf 62}, 045021 (2000); 
{\bf 62}, 085003 (2000).  
\bibitem{Charmousis}
C. Charmousis, R. Gregory and V. Rubakov,
Phys. Rev. D {\bf 62}, 067505 (2000).
\bibitem{Garriga}
J. Garriga and T. Tanaka,
Phys. Rev. Lett. {\bf 84}, 2778 (2000).


\end{references}
\end{document}